\documentclass[12pt,reqno,a4paper]{article}
\usepackage{amsmath,amsfonts,amssymb,graphicx,graphics,epsfig,pict2e,rotating}
\usepackage[english]{babel}
\numberwithin{equation}{section}

\newcommand{\be}{\begin{equation}}
\newcommand{\ee}{\end{equation}}

\begin{document}
\setcounter{page}{1}

\vspace{8mm}
\begin{center}
{\Large {\bf Finite Size Corrections for Dimers }}

\vspace{10mm}
 {\Large Alessandro Nigro\footnote{Email: Alessandro.Nigro@mi.infn.it}}\\
 [.4cm]
  
  \vspace{10mm}
 { Alessandro Nigro\\
 Dipartimento di Fisica and INFN- Sezione di Milano\\
 Universit\`a degli Studi di Milano I\\
 Via Celoria 16, I-20133 Milano, Italy\\
 Alessandro.Nigro@mi.infn.it}\\
 [.4cm]

  \end{center}

\vspace{8mm}
\centerline{{\bf{Abstract}}}
\vskip.4cm
\noindent
In this paper we derive the finite size corrections to the energy eigenvalues of the energy for 2D dimers on a square lattice. These finite size corrections, as in the case of Critical Dense Polymers, are proportional to the eigenvalues of the Local Integrals of Motion of Bazhanov Lukyanov and Zamolodchikov for central charge $c=-2$. This sheds more light on the status of the Dimer model as a conformal field theory with this value of the certral charge.
\renewcommand{\thefootnote}{\arabic{footnote}}
\setcounter{footnote}{0}
\section{Introduction}
In \cite{lieb} Lieb proposed a transfer matrix for the Dimer model, recently in \cite{dimersRR} Rasmussen and Ruelle exploited this transfer matrix description to give a partitioning of the dimer configurations into sectors which are closed under the action of the transfer matrix. In the case of open spatial boundary conditions each sector was found to be associated with a representation of the Virasoro algebra, since the partition function of a sector yields a Virasoro character in the continuum scaling limit. The characters can be organized consistently with a CFT of central charge $c=-2$.\\
The selection rules for the eigenvalues of the transfer matrix are reminiscent of those for Critical dense Polymers \cite{pera},\cite{cylpol}, since the eigenstates of the transfer matrix can be labelled by two column diagrams as well.\\
In \cite{finitesizepol} Ruelle et al. analyzed the finite size corrections to the energy levels for critical dense polymers, and found that the first correction can be described by conformal perturbation theory made with operators belonging to the tower of the identity. The analysis of finite size corrections, however had already been carried out to all orders in \cite{nigropol} albeit not by means of conformal perturbation theory. In this work it was found that each finite size correction comes with a coefficient which is proportional to the eigenvalues of the BLZ \cite{blz} Local Integrals of Motion (IOM), from which it follows immediately that the ratio of the amplitudes for finite size corrections is universal.\\
In this work we shall compute all the finite size corrections for the energy levels of the Dimer model, and show that the same result of \cite{nigropol} works in this case as well, i.e. that the finite size corrections are proportional to the eigenvalues Local IOM of \cite{blz},  as computed in \cite{nigropol}.\\
We believe that this result is interesting since it yields universal quantities of $c=-2$ CFT without using the value of the central charge as an ansatz which is checked then by consistency of the results. 
This work is organized as follows: in sections 2,3,5 we very briefly review the results of \cite{dimersRR}, whereas in section 4 we derive the exact form of the finite size corrections for Dimers.\\

\section{Dimers and Transfer Matrix}
In this section we shall briefly review the results of \cite{lieb}\cite{dimersRR} concerning the definition of the Dimers model and of a suitable transfer matrix.\\
A Dimer model on a square lattice of size $N\times M$ is a model where the lattice is covered by elementary tiles covering exactly 2 sites, the Dimers being not allowed to overlap. Each Dimer can be either orizzontal or vertical, to each dimer we associate a weight which can be rescaled so that a vertical Dimer has weight 1 and a horizontal dimer has weight $\alpha$. The partition function of such a model can be written as
\be Z_{N,M}=\sum_{\textrm{configurations}}\alpha^{h}   \ee
being $h$ the number of horizontal Dimers.\\
There is a mapping between Dimers on a square lattice and configurations of up and down arrows on the same lattice: and up arrow at site $i$  means that there is a vertical Dimer covering that site and its northern neighbour, a down arrow represents the absence of such a dimer at site $i$. A row of $N$ sites can have $2^N$ Dimer-arrow configurations.\\
It is then intuitive to think of each arrow as an element of a canonical basis $(1,0),(0,1)$ of $\mathbb{C}^2$, a row configuration then lives in the $N$-fold tensor product of $\mathbb{C}^2$ with itself.\\
We now want to build a Trasfer Matrix in terms of action of tensor products of pauli matrices, we thus define
\be \sigma_i= \mathbf{1}\otimes\mathbf{1}\otimes\ldots\otimes\sigma\otimes\ldots\otimes\mathbf{1} \ee
where $\sigma$ is in position $i$ in the tensor product and is one of the following matrices:
\be
 \sigma^x = \left( \begin{array}{cc}
0 & 1  \\
1 & 0  \end{array} \right), \qquad \sigma^- = \left( \begin{array}{cc}
0 & 0  \\
1 & 0  \end{array} \right), \qquad\sigma^+ = \left( \begin{array}{cc}
0 & 1  \\
0 & 0  \end{array} \right)
\ee
The Transfer Matrix is then defined as the product
\be T=V_3V_1 \ee
where
\be  V_1=\prod_{i=1}^{N}  \ee
and
\be V_3= \prod_{i=1}^{N-1}(\mathbf{1}+\alpha \sigma_i^-\sigma_{i+1}^-) \ee
The action of $V_1$ flips all the arrows in a given row, whereas in the action of $V_3$ each factor either lives the local arrow unchanged by acting with the term $\mathbf{1}$ or produces a horizontal dimer by reversing a pair of neighbouring up arrows by acting with $\sigma_i^-\sigma_{i+1}^-)$ and yielding a weight $\alpha$ contribution. The product $V_3V_1$ generates all possible ways of going from a prescribed set of arrow configurations by creating a horizontal dimer whenever possible and correctly handling incoming vertical dimers.\\
The partition function on a strip with periodic boundary conditions in the $M$ direction is then 
\be Z=\textrm{Tr}T^M\ee
It happens that such a partition function vanishes if both $N,M$ are odd, therefore we shall assume $M$ to be even, this also enables us to use in our analisys $T^2$ instead of $T$ itself.\\
The transfer matrix $T^2$ can be diagonalized by a Jordan-Wigner transformation followed by a Fourier transform, and its eigenvalues are of the form, for $N=2L$ even:
\be \lambda= \prod_{j=1}^L\Bigg(\sqrt{1+\alpha^2\sin^2\Big(\pi\frac{(2j-1)}{2(N+1)}\Big)}+\alpha\sin\Big(\pi\frac{(2j-1)}{2(N+1)}\Big)\Bigg )^{2(1-\epsilon_j-\mu_j)}\ee
whereas for odd $N=2L+1$
\be \lambda= \prod_{j=1}^L(\sqrt{1+\alpha^2\sin^2\Big(\frac{\pi j}{(N+1)}\Big)}+\alpha\sin\Big(\frac{\pi j}{(N+1)}\Big)\Bigg )^{2(1-\epsilon_j-\mu_j)}\ee
where $\epsilon_j, \mu_j$ independently take the values $0,1$ to such a pair $(\epsilon_j|\mu_j)$
it is possible to associate a 2-column diagram $\mathcal{D}=(j_1,\ldots,j_l|i_1,\ldots,i_r)$ there the $j$s are those for which $\epsilon_j=1$ and similarly the $i$s are those for which $\mu_i=1$.\\
The 2-column diagrams $\mathcal{D}$ will satisfy certain selection rules which we will describe in the following.\\
The energy eigenvalues are obtained from the $\lambda$s by taking the logarithm
\be E(\mathcal{D})=-\frac{1}{2}\log\lambda \ee

\section{Sectors and Variation Index}
In this section we show how to partition the space of states of the Dimer model into sectors labelled by the eigenvalues of an operator called variation index \cite{dimersRR}.\\
Following \cite{dimersRR} let us introduce the operator
\be \mathcal{V}=\frac{1}{2}\sum_{i=1}^N (-1)^{i}\sigma_i^z   \ee
This operator is digonal in the arrow basis and has eigenvalues
\be  v\in\{-\frac{N}{2},-\frac{N}{2}+1,\ldots,\frac{N}{2}\} \ee
we notice that the eigenvalues are integers for $N$ even, and half integers for $N$ odd.\\
Let $E_v$ be the vector space spanned by states having a fixed $v$, it can be shown that the generating function for the dimensions $\textrm{dim}E_v$ is:
\be \sum_{v=-\frac{N}{2}}^{\frac{N}{2}}\textrm{dim}E_v t^v=\Big(\sqrt{t}+\frac{1}{\sqrt{t}}\Big)^N  \ee
from which it follows that:
\be \textrm{dim}E_v=\textrm{dim}E_{-v}=\binom{N}{\frac{N}{2}-v}  \ee
we now want to study how the transfer matrix $T=V_3V_1$ acts on the spaces $E_v$, it is not difficult to see that $\mathcal{V}$ anticommutes with $V_1$ and commutes with $V_3$, it follows that it anticommutes with $T$, thus mapping 
\be T: \ E_v\to E_{-v} \ee
it follows that $T^2$ commutes with $\mathcal{V}$ and that each linear space $E_v$ is invariant under the action of $T^2$, whereas the $T$-invariant subspaces are of the form:
\be E_v\oplus E_{-v}\ee
we call $ E_v\oplus E_{-v}$ (and $E_0$) $T$-sectors and $E_v$ $T^2$-sectors.

\section{Finite Size Corrections}
In this section we derive the finite size corrections for the Dimer model and show that they are proportional to the eigenvalues of the integrals of motion of \cite{blz}.
\subsection{Even N}
Let the size of the system be $N=2L$, one then has that the logarithm of eigenvalues of the transfer matrix is given by
\be  E= \sum_{k\in \mathcal{D}}\textrm{arcsinh}\Big(\alpha\sin\big(\frac{\pi(2k-1)}{2(2L+1)}\big)\Big)- \sum_{k=1}^L\textrm{arcsinh}\Big(\alpha\sin\big(\frac{\pi(2k-1)}{2(2L+1)}\big)\Big)  \ee
if we now consider the following expansion
\be \textrm{arcsinh}z=\sum_{k=0}^\infty (-1)^k\frac{\big(\frac{1}{2}\big)_k}{(2k+1)k!}z^{2k+1}  \ee
we have that
\be E= \sum_{k=0}^\infty (-1)^k\frac{\big(\frac{1}{2}\big)_k}{(2k+1)k!}\alpha^{2k+1}\Big(\sum_{m\in\mathcal{D}}-\sum_{m=1}^L\Big)\sin^{2k+1}\big(\frac{\pi(2m-1)}{2(2L+1)}\big) \ee
we now have that
\be \sum_{m=1}^L \sin^{2k+1}\Big(\frac{\pi(2m-1)}{2(2L+1)}\Big)=\frac{2L+1}{\pi}\frac{(1)_k}{(\frac{3}{2})_k}-\frac{1}{2}-\sum_{l=k}^\infty\Big(\frac{\pi}{2L+1}\Big)^{2l+1}\frac{B_{2l+2}(\frac{1}{2})}{(2l+2)!}\mathcal{C}_{l,k} \ee
where
\be \sin^{2n+1}(x)=\sum_{l=n}^\infty\mathcal{C}_{l,n}\frac{x^{2l+1}}{(2l+1)!} \ee
being
\be \mathcal{C}_{l,n}=\frac{(-1)^{l+n+1}}{2^{2n+1}}\sum_{k=0}^{2n+1}(-1)^k\binom{2n+1}{k}(2(k-n)-1)^{2l+1}  \ee
it is then straightforward to obtain that
\be\begin{split}  \Big(\sum_{m\in\mathcal{D}}-\sum_{m=1}^L\Big)\sin^{2k+1}\big(\frac{\pi(2m-1)}{2(2L+1)}\big) &= -\frac{2L+1}{\pi}\frac{(1)_k}{(\frac{3}{2})_k}+\frac{1}{2}+\\
&+\sum_{l=k}^\infty\Big(\frac{\pi}{2L+1}\Big)^{2l+1}\frac{\mathcal{C}_{l,k}}{(2l+1)! 2^{2l+1}}\Big(\sum_{m\in\mathcal{D}}(2m-1)^{2l+1}+\frac{2^{2l+1}B_{2l+2}(\frac{1}{2})}{2l+2}\Big) \end{split} \ee
we now notice that
\be  \Big(\sum_{m\in\mathcal{D}}(2m-1)^{2l+1}+\frac{2^{2l+1}B_{2l+2}(\frac{1}{2})}{2l+2}\Big)=\frac{2^{l-1}}{l+1}I_{2l+1}(\mathcal{D}) \ee
where the $I_{2l+1}$ are the eigenvalues of the integrals of motion of BLZ \cite{blz} in the sectors of conformal weights $\Delta=-\frac{1}{8},\frac{3}{8},\frac{15}{8},\ldots$.\\
Now, summing up the series for the constant and divergent part gives the bulk and boundary free energy contribution, and collecting a fixed order of $2L+1$ in the sum gives the following expansion from which we read off all the finite size corrections:
\be  E= \frac{2L+1}{\pi} f_{\textrm{bulk}}(\alpha)+f_{\textrm{bou}}(\alpha)+\sum_{l=0}^\infty\Big(\frac{\pi}{2L+1}\Big)^{2l+1} \frac{P_l(\alpha)}{(2l+1)!2^{l+2}(l+1)}I_{2l+1}(\mathcal{D})  \ee
where the polynomials $P_l(\alpha)$ are defined as follows
\be P_l(\alpha)=\sum_{k=0}^l(-1)^k\frac{(\frac{1}{2})_k}{(2k+1)k!}\mathcal{C}_{l,k} \alpha^{2k+1}   \ee
the first few of these polynomials are given by
\be P_0(\alpha)= \alpha  \ee
\be P_1(\alpha)= -\alpha(1+\alpha^2)  \ee
\be P_2(\alpha)= \alpha(1+10\alpha^2+9\alpha^4)  \ee
where
\be f_{\textrm{bulk}}(\alpha)=-\int_0^{\frac{\pi}{2}} dt\ \textrm{arcsinh}(\alpha\sin t) \ee
and 
\be  f_{\textrm{bou}}(\alpha)=\frac{1}{2}\textrm{arcsinh}\alpha \ee
it can be checked that the expansion of  $f_{\textrm{bulk}}(\alpha)$ in powers of $\alpha$ is correctly reproduced by the coefficients of the expansion for the divergent part obtained here.\\
\subsection{Odd N}

Let the size of the system be $N=2L+1$, one then has that the logarithm of eigenvalues of the transfer matrix is given by
\be  E= \sum_{k\in \mathcal{D}}\textrm{arcsinh}\Big(\alpha\sin\big(\frac{\pi k}{(2L+2)}\big)\Big)- \sum_{k=1}^L\textrm{arcsinh}\Big(\alpha\sin\big(\frac{\pi k}{(2L+2)}\big)\Big)  \ee
we now have that
\be \sum_{m=1}^L \sin^{2k+1}\Big(\frac{\pi m}{(2L+2)}\Big)=\frac{2L+2}{\pi}\frac{(1)_k}{(\frac{3}{2})_k}-\frac{1}{2}-\sum_{l=k}^\infty\Big(\frac{\pi}{2L+2}\Big)^{2l+1}\frac{B_{2l+2}(0)}{(2l+2)!}\mathcal{C}_{l,k} \ee
it is then straightforward to obtain that
\be\begin{split}  \Big(\sum_{m\in\mathcal{D}}-\sum_{m=1}^L\Big)\sin^{2k+1}\big(\frac{\pi m}{(2L+2)}\big) &= -\frac{2L+2}{\pi}\frac{(1)_k}{(\frac{3}{2})_k}+\frac{1}{2}+\\
&+\sum_{l=k}^\infty\Big(\frac{\pi}{2L+2}\Big)^{2l+1}\frac{\mathcal{C}_{l,k}}{(2l+1)! 2^{2l+1}}\Big(\sum_{m\in\mathcal{D}}(2m)^{2l+1}+\frac{2^{2l+1}B_{2l+2}(0)}{2l+2}\Big) \end{split} \ee
we now notice that
\be  \Big(\sum_{m\in\mathcal{D}}(2m)^{2l+1}+\frac{2^{2l+1}B_{2l+2}(0)}{2l+2}\Big)=\frac{2^{l-1}}{l+1}I_{2l+1}(\mathcal{D}) \ee
where the $I_{2l+1}$ are the eigenvalues of the integrals of motion of BLZ in the sectors of conformal weights $\Delta=0,1,3,6,\ldots$.\\
Now, summing up the series for the constant and divergent part gives the bulk and boundary free energy contribution, and collecting a fixed order of $2L+2$ in the sum gives the following expansion from which we read off all the finite size corrections:
\be  E= \frac{2L+2}{\pi} f_{\textrm{bulk}}(\alpha)+f_{\textrm{bou}}(\alpha)+\sum_{l=0}^\infty\Big(\frac{\pi}{2L+2}\Big)^{2l+1} \frac{P_l(\alpha)}{(2l+1)!2^{l+2}(l+1)}I_{2l+1}(\mathcal{D})  \ee

\section{Characters and Selection Rules}
In this section we review the selection rules for the two-column diagrams $\mathcal{D}$ labeling the eigenstates of the integrals of motion appearing in the finite size corrections.\\
This results follow the exposition given in \cite{dimersRR}, and the conventions for the irreducible finitized characters $\textrm{ch}_{r,s}^{(N)}(q)$ of introduced in the work on critical dense polymers \cite{pera}.\\ 
\subsection{Even N}
We first of all notice that the variation index $v$ can be conveniently parametrized as:
\be  v=\sum_{j=1}^L(\mu_j-\epsilon_j)  \ee
we then recall that by neglecting the divergent and constant part the energy levels behave as:
\be E\sim\frac{\alpha\pi}{N+1}(-\frac{c}{24}-\frac{1}{8}+\sum_{j\in\mathcal{D}}(j-\frac{1}{2}))   \ee
the finitized partition function of the sector $E_v$ then reads
\be Z^N_v(q)=q^{-\frac{c}{24}-\frac{1}{8}}\sum_{\mathcal{D}\in U^{(L)}_v}q^{\sum_{j\in\mathcal{D}}(j-\frac{1}{2})}  \ee
being
\be q=e^{\frac{M\alpha\pi}{N+1}} \ee
where $U^L_v$ is the set of \emph{all} two column configurations of variation index $v$ and maximal height $L$.\\
It is possible to see that the finitized character of the sector $E_v$ is also expressible as:
\be Z^{(2L)}_v= q^{-\frac{c}{24}+\Delta_{|v|+1,2}}\bigg[{2L \atop L -v}\bigg]_{\!q}  \ee
where $\bigg[{a \atop b}\bigg]_{\!q}$ denotes the q-Binomial coefficient and 
\be  \Delta_{r,s}=\frac{(2r-s)^2-1}{8}  \ee
is the conformal  weight in the Kac table.\\
Now, by introducing the finitized irreducible characters
\be \textrm{ch}_{r,2}^{(N+1)}(q)=q^{-\frac{c}{24}+\Delta_{r,2}}\frac{1-q^{2r}}{1-q^{N+2}}\bigg[{N+1 \atop \frac{N}{2}-r+1}\bigg]_{\!q}  \ee
one can write
\be Z^{(N)}_v=\sum_{r=|v|+1, by 2}^{L or L+1}\textrm{ch}_{r,2}^{(N+1)}(q)    \ee
and in the limit $N\to\infty$:
\be Z_v(q)=\sum_{r=|v|+1, by 2}^{\infty}\textrm{ch}_{r,2}(q)    \ee
where $\textrm{ch}_{r,s}(q)$ are the irreducible characters of conformal weight $\Delta_{r,s}$ in the continuum limit. The above expression can be simplified to:
\be   Z_v(q)=\frac{q^{\frac{v^2}{2}}}{\eta(q)}  \ee
where $\eta(q)$ is the Dedekind $\eta$ function.\\

\subsection{Odd N}
Let us introduce the excess parameter $w$ defined as:
\be  w=\sum_{j=1}^L(\mu_j-\epsilon_j)  \ee
since the variation index $v$ is in this case a half integer its values are related to the excess parameter by
\be w=v\pm\frac{1}{2}  \ee
which thus has values
\be  w\in\{-L,-L+1,\ldots,L-1,L\}  \ee
again by neglecting the divergent and constant part the energy levels behave as:
\be E\sim\frac{\alpha\pi}{N+1}(-\frac{c}{24}+\sum_{j\in\mathcal{D}}j)   \ee
the finitized partition function of the sector $E_v$ then reads
\be Z^N_v(q)=q^{-\frac{c}{24}}\sum_{\mathcal{D}\in U^{(L)}_{v-\frac{1}{2}}\cup U^{(L)}_{v+\frac{1}{2}}}q^{\sum_{j\in\mathcal{D}}j}  \ee
being
\be q=e^{\frac{M\alpha\pi}{N+1}} \ee
where $U^L_w$ is the set of \emph{all} two column configurations of excess parameter $w$ and maximal height $L$.\\
It is possible to see that the finitized character of the sector $E_v$ is also expressible as:
\be Z^{(2L+1)}_v= q^{-\frac{c}{24}+\Delta_{|v|+\frac{1}{2},1}}\bigg[{2L+1 \atop L+\frac{1}{2} -v}\bigg]_{\!q}  \ee
Now, by introducing the finitized irreducible characters
\be \textrm{ch}_{r,1}^{(N+1)}(q)=q^{-\frac{c}{24}+\Delta_{r,1}}\frac{1-q^{r}}{1-q^{\frac{N+1}{2}}}\bigg[{N+1 \atop \frac{N}{2}-r+\frac{1}{2}}\bigg]_{\!q}  \ee
one can write
\be Z^{(N)}_v=\sum_{r=|v|+\frac{1}{2}}^{L+1}\textrm{ch}_{r,1}^{(N+1)}(q)    \ee
and in the limit $N\to\infty$:
\be Z_v(q)=\sum_{r=|v|+\frac{1}{2}}^{\infty}\textrm{ch}_{r,1}(q)    \ee
The above expression can again be simplified to:
\be   Z_v(q)=\frac{q^{\frac{v^2}{2}}}{\eta(q)}  \ee

\section{Acknowledgements}
The author acknowledges financial support from Fondo Sociale Europeo (Regione Lombardia), through the grant ÒDote ricercaÓ.\\


\begin{thebibliography}{}

\bibitem{lieb} E.H. Lieb, J. Math. Phys. 8 (1967) 2339.

\bibitem{dimersRR} J.Rasmussen, P.Ruelle, Refined conformal spectra in the dimer model, arxiv:1207.0385


\bibitem{pera} P.A. Pearce, J. Rasmussen,  Solvable Critical Dense Polymers, J. Stat. Mech. (2007) P02015.

\bibitem{cylpol}
P.Pearce, J.Rasmussen, S.P.Villani, Solvable Critical Dense Polymers on the Cylinder, J.Stat.Mech.1002:P02010,2010 


\bibitem{finitesizepol}N.S. Izmailian, P.Ruelle,C.K. Hu, Finite size corrections for logarithmic representations in critical dense polymers, Physics Letters B 711 (2012) 71-75

\bibitem{nigropol} 

A. Nigro, Integrals of Motion for Critical Dense Polymers and Symplectic Fermions, J. Stat. Mech. (2009) P10007

\bibitem{blz}
VV. Bazanov, S.L. Lukyanov, A.B. Zamolodchikov, Commun.Math.Phys. 177 (1996) 381-398 










\bibitem{tba}
P.Grinza, G.Feverati, Integrals of Motion from TBA and lattice-conformal dictionary,  Nucl.Phys. B702 (2004) 495-515 




\end{thebibliography}
\end{document}